\documentclass{article} 
\usepackage[preprint]{colm2026_conference}

\ifdefined\pdfobjcompresslevel
\fi

\usepackage{microtype}
\usepackage{hyperref}
\usepackage{url}
\usepackage{booktabs}
\usepackage{graphicx}
\usepackage{algorithm}
\usepackage{algpseudocode}
\usepackage{multirow}
\usepackage{array}
\usepackage{tcolorbox}
\usepackage[table]{xcolor}
\usepackage{colortbl}
\usepackage{lineno}
\usepackage{amsmath}
\usepackage{amssymb}
\usepackage{xcolor}
\usepackage{longtable}
\usepackage{enumitem}
\usepackage{newtxtext,newtxmath}
\usepackage{titletoc}

\definecolor{darkblue}{rgb}{0, 0, 0.5}
\hypersetup{
    colorlinks=true,
    citecolor=darkblue,
    linkcolor=darkblue,
    urlcolor=darkblue
}

\title{MA-SAPO: Multi-Agent Reasoning for Score-Aware Prompt Optimization}

\author{%
Juhyeon Lee$^{1}$\thanks{Equal contribution.}\quad
Wonduk Seo$^{2*}$\thanks{This work was done while Wonduk Seo was at Enhans.}\quad
Junseo Koh$^{1}$\quad Wonseok Choi$^{1}$\quad Hyunjin An$^{2}$\\
\textbf{Jian Park$^{3}$\quad
Seunghyun Lee$^{2}$\quad
Haihua Chen$^{4\ddagger}$\quad
Yi Bu$^{1}$\thanks{Corresponding author(s): Haihua Chen, Yi Bu.}}\\
$^1$\it{Peking University} $^2$\it{Enhans} $^3$\it{Fudan University} $^4$\it{University of North Texas}\\
\texttt{\{leejuhyeon,junseokoh,choiwonseok\}@stu.pku.edu.cn, wonduk@enhans.ai}\\
\texttt{haihua.chen@unt.edu, buyi@pku.edu.cn}
}

\begin{document}

\ifcolmsubmission
\linenumbers
\fi

\maketitle
\begin{abstract}
Prompt optimization has become a practical way to improve the performance of Large Language Models (LLMs) without retraining. However, most existing frameworks treat evaluation as a black box, relying solely on outcome scores without explaining why prompts succeed or fail. Moreover, they involve repetitive trial-and-error refinements that remain implicit, offering limited interpretability or actionable guidance for systematic improvement. 
In this paper, we propose \textbf{MA-SAPO}: a new \textbf{\underline{M}}ulti-\textbf{\underline{A}}gent Reasoning for \textbf{\underline{S}}core \textbf{\underline{A}}ware \textbf{\underline{P}}rompt \textbf{\underline{O}}ptimization framework that links evaluation outcomes directly to targeted refinements. Specifically, in the \emph{Training Phase}, multiple agents interpret evaluation scores, diagnose weaknesses, and generate concrete revision directives, which are stored as reusable reasoning assets. In the \emph{Test Phase}, an analyzer agent retrieves relevant exemplars and assets for a new prompt, and a refiner agent applies evidence-based edits to improve the prompt and its response.By grounding optimization in structured reasoning, \textbf{MA-SAPO} ensures edits are interpretable, auditable, and controllable. Experiments on the \emph{\emph{HelpSteer1}/2} benchmarks show that our framework consistently outperforms single-pass prompting, retrieval-augmented generation, and prior multi-agent methods across multiple evaluation metrics.
\end{abstract}

\section{Introduction}

Large Language Models (LLMs) have emerged as powerful tools capable of tackling a wide range of tasks, from reasoning and summarization to complex dialogue and code generation~\citep{brown2020language,chowdhery2023palm,touvron2023llama}. However, their performance remains highly sensitive to the wording  and examples embedded within prompts~\citep{pryzantautomatic,he2024does,cao2024worst,razavi2025benchmarking}. Consequently, prompt optimization has rapidly emerged as a practical alternative to costly retraining or parameter updates for improving model behavior.

Early research explored single-pass prompt optimization that effectively treats the LLM as its own optimizer. Approaches such as Chain-of-Thought (CoT) prompting~\citep{wei2022chain}, role conditioning via carefully crafted system instructions~\citep{wu2023large,kong2024better}, and structured reasoning schemes including Tree-of-Thought (ToT)~\citep{yao2023tree}, Graph-of-Thought (GoT)~\citep{besta2024graph}, and step-back prompting~\citep{zhengtake} demonstrated that encoding explicit reasoning structure or role signals in the prompt can materially improve downstream performance. Nevertheless, these methods are largely self-contained, relying on a single model instance to revise prompts without leveraging systematic feedback across iterations or external resources.

As LLMs have advanced, multi-agent frameworks have emerged to broaden viewpoints and coordinate specialized roles in prompt optimization~\citep{zhoumulti,zhang2025mars,han2025mapgd}. By delegating subtasks such as decomposition, evaluation, and refinement to distinct agents, these systems extend beyond single-pass rewriting into more structured pipelines. In parallel, optimization-theoretic approaches~\citep{li2025survey,cui2025automatic} and label-efficient schemes~\citep{xiang2025self,dong2025model} aim to curb reliance on expensive feedback while scaling across tasks.

Despite these advances, several common limitations remain: (i) evaluation is often collapsed to scalar scores, obscuring why prompts succeed or fail; (ii) refinement remains largely trial-and-error, incurring high computational cost with limited transparency; (iii) reasoning, when used, is typically implicit rather than distilled into auditable artifacts that drive targeted edits; and (iv) pipelines prioritize score gains over interpretability and controllability, which hinders practitioners from analyzing trade-offs and making precise adjustments in real-world applications.

To address these limitations, we introduce \textbf{MA-SAPO}, a sequential reasoning framework for prompt optimization that explicitly maps metric outcomes to actionable edits via structured reasoning artifacts. \textbf{MA-SAPO} has two phases. In the \emph{training phase}, three agents convert annotated scores into reusable assets: (1) a \textbf{Metric Explainer} interprets evaluation dimensions, (2) a \textbf{Diagnostician} identifies error sources and trade-offs, and (3) an \textbf{Action Synthesizer} produces concrete edit directives. In the \emph{test phase}, a retrieval-augmented pipeline applies these assets: an \textbf{Analyzer} contrasts the current prompt with retrieved exemplars to surface improvement opportunities, and a \textbf{Refiner} performs targeted edits grounded in diagnostic evidence, yielding interpretable and controllable optimization.

Experiments on \textit{HelpSteer1/2}~\citep{wang2024helpsteer1,wang2024helpsteer2} show that \textbf{MA-SAPO} consistently surpasses single-pass prompting, retrieval-augmented generation, and prior multi-agent baselines, underscoring structured reasoning as an effective bridge between evaluation and optimization. Our contributions are:
\begin{itemize}
    \item A score-aware multi-agent training pipeline that distills evaluation outcomes into \textbf{reusable, semi-structured reasoning assets} for explanation, diagnosis, and edit directives.
    \item \textbf{MA-SAPO}, a modular and compute-efficient framework that delivers \textbf{interpretable, auditable, and controllable} gains while substantially \textbf{reducing token and API-call budgets}.
    \item Extensive results on HelpSteer1/2 demonstrating \textbf{consistent multi-metric improvements} and outperforming six state-of-the-art baselines in both effectiveness and efficiency.
\end{itemize}

\section{Related work}

\subsection{Single-pass prompt optimization}

Prompt optimization has rapidly evolved as a practical means of improving the performance of LLMs without retraining~\citep{wei2022chain,pryzantautomatic,sun2023autohint,li2025test}. Early studies highlighted the centrality of prompts as control variables for steering model behavior. Specifically, few-shot prompting~\citep{brown2020language} demonstrated that models could adapt to unseen tasks using in-context examples alone, while retrieval-augmented generation (RAG)~\citep{lewis2020retrieval,lee2025better} injected external knowledge to improve factuality and adaptability. These two paradigms, context construction and evidence injection, laid the foundation for single-pass prompt optimization methods. 
Subsequent research explored reasoning-augmented prompting. Chain-of-Thought (CoT)~\citep{wei2022chain} introduced intermediate reasoning steps. Building on this idea, Tree-of-Thought (ToT)~\citep{yao2023tree} generalized linear reasoning chains into branching structures with self-evaluation via lookahead and backtracking. Other structured strategies, such as Skeleton-of-Thought (SoT)~\citep{ning2023skeleton}, separate planning from execution to parallelize generation, while Graph-of-Thought (GoT)~\citep{besta2024graph} organizes reasoning trajectories into graph topologies. Evolutionary and heuristic single-pass approaches have also been explored. For instance, PromptBreeder~\citep{fernando2023promptbreeder} evolves a population of prompts through mutation and selection, while PromptWizard~\citep{agarwal2024promptwizard} refines prompts via feedback-driven critique-and-synthesis loops. 

\subsection{Multi-agent based prompt optimization}

Beyond single-pass methods, multi-agent systems have emerged as a powerful paradigm for prompt optimization~\citep{du2024improving,liutowards,chen2024comm,licoevol,seo2026new}. Early agentic frameworks such as Self-Ask~\citep{press2023measuring} introduced decomposition by generating sub-questions and integrating external retrieval, laying the groundwork for cooperative prompting strategies. Multi-Agent Debate (MAD)~\citep{du2024improving} advanced this idea by allowing agents to iteratively present and critique answers in a roundtable format, thereby improving factual grounding and reasoning quality. Reflection-based frameworks~\citep{shinn2023reflexion,khan2024debating} further enhanced adaptability by storing feedback as memory for subsequent optimization attempts. 
Building on this foundation, recent systems formalized agent roles and workflows to optimize prompts more systematically~\citep{li2023camel}. Subsequent work explicitly targeted prompt optimization as the objective. For instance, MASS~\citep{zhoumulti} searches over agent topologies while performing block-level prompt refinement, followed by global optimization. Additionally, MARS~\citep{zhang2025mars} leverages a Planner and Teacher--Critic--Student dialogue to iteratively propose, critique, and revise prompts in structured rounds. MAPGD~\citep{han2025mapgd} further integrates pseudo-gradient feedback with bandit-style exploration, offering sample-efficient optimization through agent collaboration. However, most existing frameworks treat evaluation as a black box with limited interpretability and rely on costly trial-and-error refinements. \textbf{MA-SAPO} addresses this by grounding optimization in score-aware reasoning assets that link evaluation outcomes to actionable, interpretable, and auditable refinements.

\section{Methodology}

\begin{figure}[h]
    \centering
    \includegraphics[width=\linewidth]{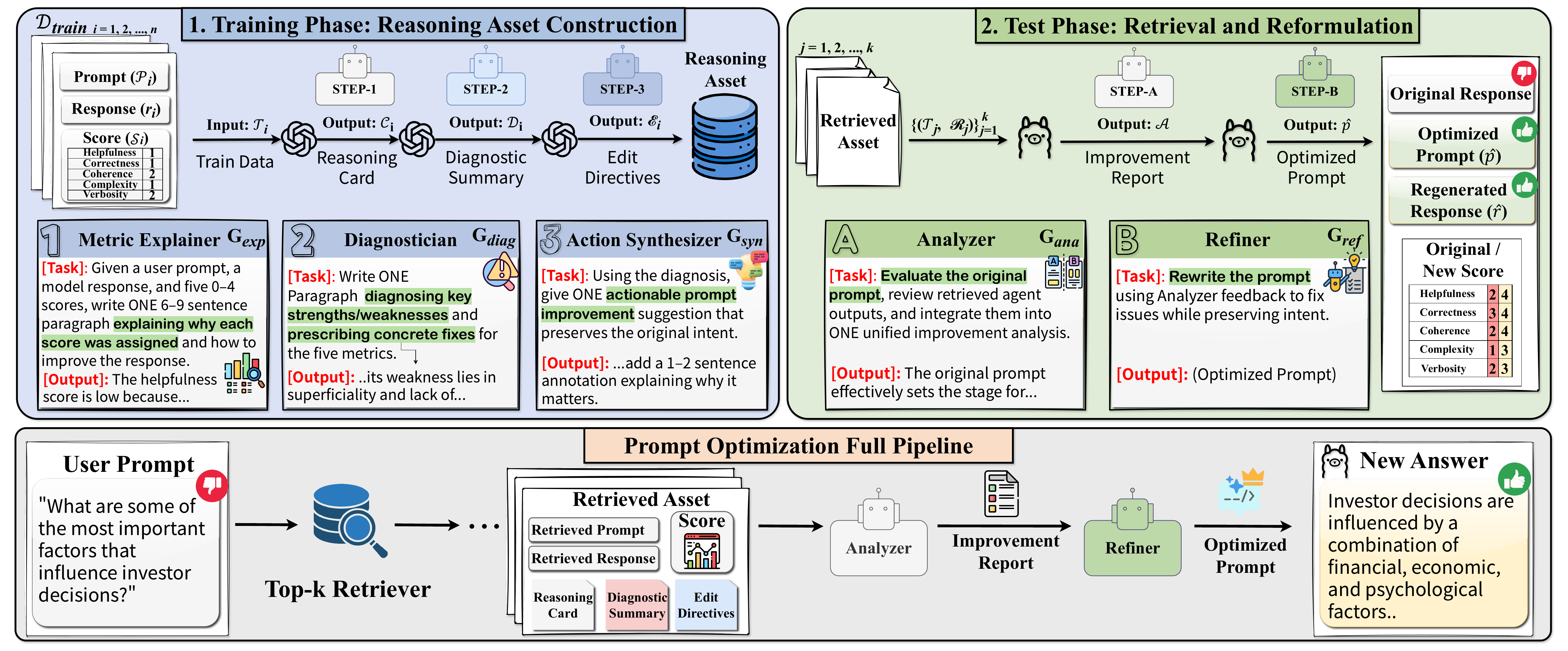}
    \caption{MA-SAPO pipeline overview. Three training agents convert annotated 
    prompt--response pairs into reusable Reasoning Assets $\mathcal{R}_i = 
    (\mathcal{C}_i, \mathcal{D}_i, \mathcal{E}_i)$. At test time, retrieved assets 
    guide the Analyzer $G_{\mathrm{ana}}$ and Refiner $G_{\mathrm{ref}}$ to produce 
    an optimized prompt $\hat{p}$ with higher scores across all five evaluation dimensions.}
    \label{fig:framework}
\end{figure}

\subsection{Preliminaries}

Let the training dataset be denoted as
\[
\mathcal{D}_{train} = \{\tau_i\}_{i=1}^N, \quad
\tau_i = (p_i, r_i, \mathcal{S}_i),
\]
where each entry $\tau_i$ consists of a prompt $p_i$, its associated response $r_i$, and a set of scores
$\mathcal{S}_i = \{s_{help}, s_{corr}, s_{coh}, s_{comp}, s_{verb}\}$, corresponding to the dimensions of \emph{helpfulness}, \emph{correctness}, \emph{coherence}, \emph{complexity}, and \emph{verbosity}.
The test set is denoted as
\[
\mathcal{D}_{test} = \{p_j\}_{j=1}^M,
\]
where only prompts are available, and corresponding responses are generated during evaluation. Our objective is to generate structured reasoning assets from $\mathcal{D}_{train}$ that can be retrieved to optimize new prompts and their corresponding responses. Thus, the training phase serves as \emph{data generation}, while the test phase operates as a \emph{retrieval-augmented optimization} pipeline.

Throughout this framework, we employ several specialized agents denoted as $G$. Each agent is designed with a distinct functional role: $G_{\text{exp}}$ explains metric-level outcomes, $G_{\text{diag}}$ diagnoses weaknesses and trade-offs, $G_{\text{syn}}$ synthesizes actionable directives, $G_{\text{ana}}$ analyzes retrieved examples, and $G_{\text{ref}}$ refines prompts into optimized versions. These agents are executed in sequential order within their respective phases, and their outputs are explicitly stored for later retrieval and reasoning. The overview of the \textbf{MA-SAPO} framework is shown in Figure~\ref{fig:framework}.

\subsection{Training phase: reasoning asset construction}

In the training phase, three agents are executed sequentially for each $\tau_i \in \mathcal{D}_{train}$. All three agents operate on the same triplet $\tau_i=(p_i, r_i, \mathcal{S}_i)$, ensuring that their reasoning is consistently grounded in both the input and output of the original instance.

\subsubsection{Metric explainer agent \texorpdfstring{$G_{\text{exp}}$}{Gexp}}

The Metric Explainer Agent generates natural language justifications for the assigned scores.
Formally, given $\tau_i$, the agent produces a reasoning card $\mathcal{C}_i$:
\begin{equation}
\mathcal{C}_i = G_{\text{exp}}(\tau_i, P_{\text{exp}}),
\end{equation}
where $P_{\text{exp}}$ is the system prompt guiding explanation.
Each card explicitly outlines the reason why $\tau_i$ received its annotated scores.
The reasoning card $\mathcal{C}_i$ then serves as input to the next agent, together with $\tau_i$.

\subsubsection{Diagnostician agent \texorpdfstring{$G_{\text{diag}}$}{Gdiag}}

The Diagnostician Agent operates on $\tau_i$ and the reasoning card $\mathcal{C}_i$.
It extends $\mathcal{C}_i$ by analyzing metric-level weaknesses and trade-offs.
$G_{\text{diag}}$ produces a diagnostic summary $\mathcal{D}_i$ that identifies:
(1) the key causes of low-scoring dimensions, and
(2) trade-offs across metrics (e.g., verbosity vs.\ coherence).
\begin{equation}
\mathcal{D}_i = G_{\text{diag}}(\tau_i, \mathcal{C}_i, P_{\text{diag}}).
\end{equation}
The diagnostic summary $\mathcal{D}_i$ is then passed to the next stage, along with $\tau_i$.
    
\subsubsection{Action synthesizer agent \texorpdfstring{$G_{\text{syn}}$}{Gsyn}}

The Action Synthesizer Agent receives $\tau_i$ together with $\mathcal{C}_i$ and $\mathcal{D}_i$.
It converts these enriched insights into actionable edit directives (EDs). Each directive corresponds to a concrete modification strategy for improving $\tau_i$:
\begin{equation}
\mathcal{E}_i = G_{\text{syn}}(\tau_i, \mathcal{C}_i, \mathcal{D}_i, P_{\text{syn}}),
\end{equation}
where $\mathcal{E}_i = \{e_1, e_2, \dots, e_m\}$ is a set of recommended edits.

\paragraph{Definition (reasoning assets).}

For notational convenience, we define the collection of training outputs for instance $i$ as:
\[
\mathcal{R}_i = (\mathcal{C}_i, \mathcal{D}_i, \mathcal{E}_i).
\]
As a result, $\mathcal{R}_i$ is stored as reasoning assets aligned with $\tau_i$, forming the retrieval corpus for the test phase. Importantly, these reasoning assets are designed as semi-structured text, which makes them machine-parseable and directly usable by downstream agents in the test phase. For instance, edit directives often follow a pattern such as “add a constraint to ...”, which the refiner agent can reliably identify and execute.

\subsection{Test phase: retrieval and reformulation}

In the test phase, optimization is performed via retrieval-augmented generation.
Given a new prompt $p_{\text{test}}$, we adopt a sparse lexical retriever that ranks training prompts based on term-frequency and inverse document-frequency statistics.
Each training prompt $p_j$ is represented in the sparse lexical space, and a relevance score is computed between $p_{\text{test}}$ and $p_j$ according to token overlap weighted by discriminative importance.
The top-$k$ most relevant prompt-response pairs are then retrieved along with their corresponding reasoning assets:
\[
\{(\tau_j, \mathcal{R}_j)\}_{j=1}^k.
\]

\subsubsection{Analyzer agent \texorpdfstring{$G_{\text{ana}}$}{Gana}}

The Analyzer Agent compares $p_{\text{test}}$ against retrieved examples to identify improvement points. It outputs an improvement report $\mathcal{A}$:
\begin{equation}
\mathcal{A} = G_{\text{ana}}(p_{\text{test}}, \{\tau_i, \mathcal{R}_j\}_{j=1}^k, P_{\text{ana}}).
\end{equation}
The Analyzer Agent transforms raw retrieval into structured insights rather than merely transferring edits from similar prompts. The agent highlights concrete weaknesses and improvement opportunities by contrasting the test prompt with retrieved reasoning assets.

\subsubsection{Refiner agent \texorpdfstring{$G_{\text{ref}}$}{Gref}}

The Refiner Agent regenerates an optimized prompt $\hat{p}$ by incorporating the improvement report:
\begin{equation}
\hat{p} = G_{\text{ref}}(p_{\text{test}}, \mathcal{A}, P_{\text{ref}}).
\end{equation}
It operationalizes the analyzer’s insights into a concrete optimized prompt. Compared to a direct rewriting approach, which may introduce irrelevant or unjustified changes, the refiner explicitly conditions on the improvement report, producing refinements that are focused, justifiable, and consistent with diagnostic evidence. The final optimized prompt $\hat{p}$ is then used to generate a corresponding optimized response $\hat{r}$.\footnote{The detailed 
prompt templates and specific role descriptions for all agents are 
provided in Appendix~\ref{prompt_design}.}

\section{Experimental design and setup}

\subsection{Datasets}
We conduct our experiments on the HelpSteer family of datasets~\citep{wang2024helpsteer1,wang2024helpsteer2}, 
which provide human-annotated prompt-response pairs scored across five quality dimensions on a 0--4 scale.\footnote{Detailed descriptions of the five annotation dimensions are provided in Table~\ref{tab:dataset_attributes} in Appendix~\ref{app:dataset}.} 
Both datasets share the same annotation schema but differ in size and design, as summarized in 
Table~\ref{tab:dataset_stats}. We build our retrieval corpus exclusively from the HelpSteer2 training set, 
as it offers higher-quality annotations and broader prompt diversity, with approximately 29\% of samples 
consisting of multi-turn interactions. Furthermore, its preference annotations between responses provide 
finer-grained evidence for our Analyzer and Refiner agents, and it serves as the current benchmark standard 
for training and evaluating reward models. Evaluation is conducted on the validation splits of 
both HelpSteer1 and HelpSteer2.

\begin{table}[ht]
\centering
\small
\begin{tabular}{lcccc}
\toprule
\textbf{Dataset} & \textbf{Train} & \textbf{Val} & \textbf{Multi-turn} & \textbf{Preference Ann.} \\
\midrule
HelpSteer1 & 35.3k & 1.79k & \texttimes & \texttimes \\
HelpSteer2 & 20.3k & 1.04k & \checkmark & \checkmark \\
\bottomrule
\end{tabular}
\vspace{6pt}
\caption{Statistics of the HelpSteer1/2 datasets. Both share the same annotation schema of five quality 
dimensions scored on a 0--4 scale.}
\label{tab:dataset_stats}
\end{table}

\subsection{Models used}
\paragraph{Reasoning--asset generation (corpus construction).}
To construct the retrieval corpus, we employ the three reasoning agents 
($G_{\text{exp}}$, $G_{\text{diag}}$, $G_{\text{syn}}$) using the 
\textit{o4-mini reasoning} model from OpenAI~\citep{achiam2023gpt}.
This model is used exclusively for \emph{reasoning asset construction} rather than downstream inference. All generations follow default API hyperparameters.
For each training instance $\tau_i$, it produces a reasoning card $\mathcal{C}_i$,
a diagnostic summary $\mathcal{D}_i$, and a set of edit directives $\mathcal{E}_i$,
which together form the stored reasoning assets $\mathcal{R}_i$.

\paragraph{Generation models for evaluation.}
For downstream evaluation, we adopt a representative API model \textit{GPT-4o}~\citep{achiam2023gpt}, and an open-source model \textit{Llama3-8B-Instruct}~\citep{dubey2024llama}
as the backbone models.
These models are responsible for generating (i) optimized prompts $\hat{p}$ from the Refiner Agent
and (ii) the corresponding optimized responses $\hat{r}$ conditioned on $\hat{p}$.
Both models are configured with temperature $=0$ to ensure deterministic decoding,
and all other hyperparameters remain at their defaults.

\paragraph{Judge model.}
We employ the \textit{ArmoRM-Llama3-8B-v0.1} reward model~\citep{ArmoRM} as an interpretable multi-objective evaluator ($>90\%$ benchmark accuracy), which automatically scores response quality. For any prompt-response pair, the judge model returns five HelpSteer-aligned scores:
\[
\mathcal{M}=\{\text{help}, \text{corr}, \text{coh}, \text{comp}, \text{verb}\},
\quad s_m \in [0,4].
\]

\paragraph{Retrieval.}
We adopt a sparse retriever based on BM25~\citep{robertson2009probabilistic} over prompt text.
For each test prompt $p_{\text{test}}$, we retrieve the top-$k$ training prompts with $k=3$, which we identify as the optimal setting for retrieval\footnote{A detailed analysis of $k$ values and their effects on performance is provided in Section~\ref{sec:topk-retrieval-variations}.}
and attach their paired responses and reasoning assets:
\begin{equation}
\{(p_j, r_j, \mathcal{R}_j)\}_{j=1}^{k=3}.
\end{equation}

\subsection{Evaluation}
\paragraph{Evaluation metrics.}
To assess the effectiveness of our framework, we evaluate each candidate pair 
$(p,r)$ is evaluated across the five HelpSteer dimensions 
$\mathcal{M} = \{\text{help}, \text{corr}, \text{coh}, \text{comp}, 
\text{verb}\}$, where each raw score $s_m \in [0,4]$. We normalize 
scores by dividing by $4$, yielding $\tilde{s}_m = s_m/4 \in [0,1]$.
The overall quality score is then defined as the average over all 
metrics:
\begin{equation}
\mathrm{Score}(p,r) = \tfrac{1}{|\mathcal{M}|} \sum_{m \in \mathcal{M}} \tfrac{s_m}{4}.
\end{equation}
This produces a single composite value in the range $[0,1]$, 
reflecting the overall response quality. We additionally 
complement automatic metrics with a human evaluation involving 17 expert annotators, which examines (H1) the 
usefulness, accuracy, and consistency of the reasoning and (H2) 
whether optimized prompts preserve the original intent.\footnote{The results of this human evaluation are presented in Section~\ref{sec:human-eval}, with detailed protocols and per-dimension scores in Appendix~\ref{app:qualitative}.}

\subsection{Baselines}

We compare our method with six baselines in three settings: single-pass prompt optimization without retrieval (\textit{Direct Generation}, \textit{CoT}~\citep{wei2022chain}, and \textit{Role Assignment}~\citep{wu2023large}), retrieval-augmented optimization (\textit{RAG}~\citep{lewis2020retrieval}), and multi-agent optimization (\textit{MAD}~\citep{du2024improving} and \textit{MARS}~\citep{zhang2025mars}). For all methods, the model first generates an optimized prompt from the input, and then produces the final response from that optimized prompt using the same backbone LLM. We use \textit{GPT-4o} and \textit{Llama3-8B-Instruct} for all baselines, with temperature set to $0$ and other decoding settings kept at their defaults. For \textsc{RAG}, we retrieve the top-$10$ training examples using BM25. For \textsc{MAD}, we use $2$ agents and $3$ debate rounds. For \textsc{MARS}, we use $3$--$4$ planning steps and $5$ refinement iterations.\footnote{Baseline details are provided in Appendix~\ref{app:baseline_app}.}

\section{Experimental results and analysis}

\subsection{Main results}

\begin{table}[t]
\centering
\scriptsize
\setlength{\tabcolsep}{3.2pt}
\renewcommand{\arraystretch}{1.05}
\resizebox{\linewidth}{!}{%
\begin{tabular}{l|cccccc|cccccc}
\toprule
\multirow{2}{*}{\textbf{Methods}} 
 & \multicolumn{6}{c|}{\textbf{GPT-4o}} & \multicolumn{6}{c}{\textbf{Llama3-8B}} \\
 & \textbf{Help} & \textbf{Corr} & \textbf{Coh} & \textbf{Comp} & \textbf{Verb} & \textbf{Avg} 
 & \textbf{Help} & \textbf{Corr} & \textbf{Coh} & \textbf{Comp} & \textbf{Verb} & \textbf{Avg} \\
\midrule

\rowcolor{gray!15}
\multicolumn{13}{l}{\textbf{\emph{HelpSteer1}}} \\
Direct Generation & 0.3216 & 0.3866 & 0.7583 & 0.2951 & 0.4215 & 0.4366 & 0.2549 & 0.3247 & 0.7054 & 0.2702 & 0.4062 & 0.3927 \\
RAG (sparse, $k{=}10$)~(\citeyear{lewis2020retrieval}) & 0.3751 & 0.4402 & 0.7871 & 0.3116 & 0.4779 & 0.4784 & 0.2930 & 0.3594 & 0.7452 & 0.3007 & 0.4377 & 0.4272 \\
Chain-of-Thought (CoT)~(\citeyear{wei2022chain}) & 0.3223 & 0.3876 & 0.7595 & 0.2935 & 0.4192 & 0.4364 & 0.2359 & 0.3077 & 0.6906 & 0.2627 & 0.3967 & 0.3787 \\
Role Assignment~(\citeyear{shanahan2023role}) & 0.3988 & 0.4679 & \underline{0.8024} & \underline{0.3427} & 0.5008 & 0.5025 & 0.2887 & 0.3516 & 0.7287 & 0.3164 & 0.4517 & 0.4274 \\
MAD~(\citeyear{du2024improving}) & 0.3774 & 0.4764 & 0.7954 & 0.3210 & \underline{0.5248} & 0.4990 & 0.3774 & \textbf{0.5049} & \underline{0.8067} & \underline{0.4084} & \underline{0.6708} & \underline{0.5531} \\
MARS~(\citeyear{zhang2025mars}) & \underline{0.4159} & \underline{0.4876} & 0.7963 & 0.3293 & 0.5188 & \underline{0.5096} & \underline{0.3901} & 0.4591 & 0.7745 & 0.3512 & 0.5801 & 0.5110 \\
\textbf{MA-SAPO (Ours)} & \textbf{0.5183} & \textbf{0.6260} & \textbf{0.8614} & \textbf{0.5013} & \textbf{0.7363} & \textbf{0.6486} & \textbf{0.4110} & \underline{0.4868} & \textbf{0.8326} & \textbf{0.4720} & \textbf{0.8433} & \textbf{0.6091} \\

\midrule
\rowcolor{gray!15}
\multicolumn{13}{l}{\textbf{\emph{HelpSteer2}}} \\
Direct Generation & 0.3616 & 0.4700 & 0.7723 & 0.3280 & 0.4913 & 0.4846 & 0.2616 & 0.3636 & 0.7078 & 0.2942 & 0.4421 & 0.4139 \\
RAG (sparse, $k{=}10$)~(\citeyear{lewis2020retrieval}) & \underline{0.4903} & \underline{0.5745} & \textbf{0.8642} & \underline{0.4161} & 0.6567 & \underline{0.6003} & 0.3990 & 0.4711 & 0.7989 & 0.3722 & 0.5814 & 0.5245 \\
Chain-of-Thought (CoT)~(\citeyear{wei2022chain}) & 0.2981 & 0.3958 & 0.7169 & 0.2888 & 0.4709 & 0.4341 & 0.1600 & 0.2545 & 0.6291 & 0.2421 & 0.3812 & 0.3334 \\
Role Assignment~(\citeyear{shanahan2023role}) & 0.4400 & 0.5175 & 0.8221 & 0.4025 & 0.5992 & 0.5563 & 0.2555 & 0.3303 & 0.7050 & 0.3267 & 0.4441 & 0.4123 \\
MAD~(\citeyear{du2024improving}) & 0.4167 & 0.5049 & 0.8067 & 0.4084 & \underline{0.6708} & 0.5615 & 0.3971 & 0.4532 & 0.7898 & \underline{0.3998} & \underline{0.6439} & 0.5368 \\
MARS~(\citeyear{zhang2025mars}) & 0.4791 & 0.5569 & 0.8268 & 0.4095 & 0.6234 & 0.5791 & \textbf{0.4296} & \textbf{0.5019} & \underline{0.7994} & 0.3957 & 0.6181 & \underline{0.5482} \\
\textbf{MA-SAPO (Ours)} & \textbf{0.5072} & \textbf{0.6038} & \underline{0.8527} & \textbf{0.5244} & \textbf{0.7570} & \textbf{0.6490} & \underline{0.4005} & \underline{0.4754} & \textbf{0.8294} & \textbf{0.4833} & \textbf{0.8441} & \textbf{0.6065} \\

\bottomrule
\end{tabular}%
}
\caption{Main results on \emph{HelpSteer1} and \emph{HelpSteer2}. Columns report the five HelpSteer metrics (Help, Corr, Coh, Comp, Verb) normalized to $[0,1]$ and their mean (\textbf{Avg}). Methods are grouped into (i) single-pass prompting, (ii) retrieval-augmented generation (no reasoning assets), and (iii) multi-agent frameworks. For each dataset and backbone block, the best value in each metric column is shown in \textbf{bold}, and the second-best is \underline{underlined}.}
\label{tab:main_helpsteer}
\end{table}

In our main experiments, we compare \textbf{MA-SAPO} with three groups of baselines: (1) single-pass prompting methods (\textit{Direct Generation}, \textit{CoT}, and \textit{Role Assignment}), (2) retrieval-augmented generation (\textit{RAG}) without reasoning assets, and (3) multi-agent frameworks (\textit{MAD} and \textit{MARS}). As shown in Table~\ref{tab:main_helpsteer}, \textbf{MA-SAPO} consistently achieves the best results across both HelpSteer1 and HelpSteer2, under both GPT-4o and Llama3-8B. Single-pass methods remain limited because they revise prompts in isolation, without diagnostic grounding, which often leads to brittle or inconsistent improvements across metrics. Although Role Assignment introduces more explicit guidance, it still depends heavily on predefined roles. RAG benefits from retrieved exemplars, but without explicit reasoning about why those examples are useful, its refinements remain heuristic and less reliable.

\begin{figure}[t]
    \centering
    \includegraphics[width=0.8\linewidth]{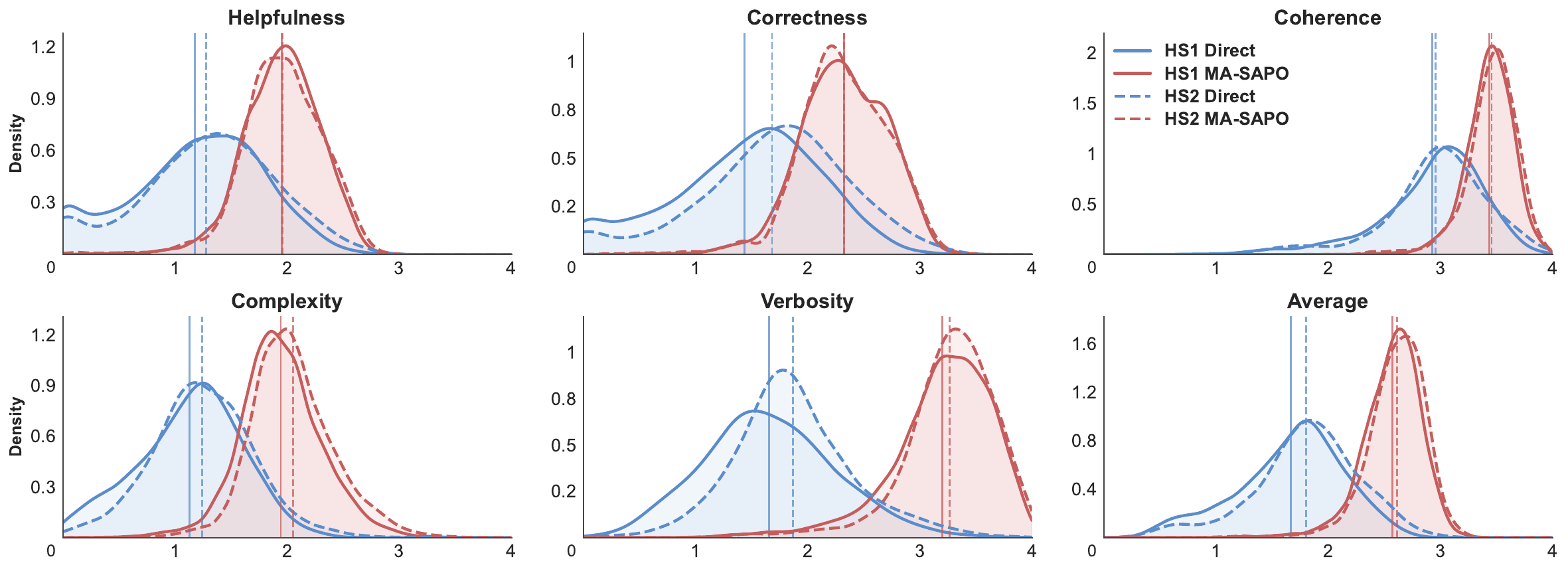}
    \caption{KDE score distributions of Direct Generation and MA-SAPO across five HelpSteer dimensions and their average, aggregated over GPT-4o and Llama3-8B. Solid lines denote HelpSteer1 and dashed lines denote HelpSteer2. Vertical lines indicate the mean of each distribution. MA-SAPO consistently shifts the score distribution toward higher values.\protect\footnotemark}
    \label{fig:kde_combined}
\end{figure}
\footnotetext{Distributions for GPT-4o and Llama3-8B are provided in Appendix~\ref{app:per_backbone_kde}.}

\begin{table*}[t]
\centering
\scriptsize
\setlength{\tabcolsep}{3.4pt}
\renewcommand{\arraystretch}{1.08}
\resizebox{\textwidth}{!}{%
\begin{tabular}{l|cccccc|cccccc}
\toprule
\multirow{2.5}{*}{\textbf{Methods}}
& \multicolumn{6}{c|}{\textbf{GPT-4o}} 
& \multicolumn{6}{c}{\textbf{Llama3-8B}} \\
\cmidrule(lr){2-7}\cmidrule(lr){8-13}
& \textbf{Help} & \textbf{Corr} & \textbf{Coh} & \textbf{Comp} & \textbf{Verb} & \textbf{Avg}
& \textbf{Help} & \textbf{Corr} & \textbf{Coh} & \textbf{Comp} & \textbf{Verb} & \textbf{Avg} \\
\midrule

\rowcolor{gray!15}
\multicolumn{13}{l}{\textbf{\emph{HelpSteer1}}} \\
$k=1$ & 0.5182 & 0.6247 & 0.8543 & 0.4943 & \underline{0.7349} & 0.6453 & \underline{0.4045} & 0.4769 & 0.8268 & \textbf{0.4729} & \textbf{0.8446} & \underline{0.6051} \\
$k=2$ & \textbf{0.5211} & \textbf{0.6287} & 0.8591 & 0.4960 & 0.7318 & \underline{0.6473} & 0.4002 & 0.4747 & 0.8236 & \underline{0.4726} & 0.8384 & 0.6019 \\
$k=4$ & \underline{0.5204} & \underline{0.6266} & 0.8606 & \underline{0.4965} & \textbf{0.7369} & \textbf{0.6482} & 0.4028 & 0.4778 & 0.8272 & 0.4708 & 0.8389 & 0.6035 \\
Test Agents Combination & 0.4672 & 0.5415 & \textbf{0.8619} & 0.3670 & 0.6403 & 0.5756 & \textbf{0.4649} & \textbf{0.5203} & \underline{0.8323} & 0.4363 & 0.7249 & 0.5957 \\
\textbf{MA-SAPO ($k$=3)} & 0.5183 & 0.6260 & \underline{0.8614} & \textbf{0.5013} & 0.7363 & 0.6486 & 0.4110 & \underline{0.4868} & \textbf{0.8326} & 0.4720 & \underline{0.8433} & \textbf{0.6091} \\
\midrule

\rowcolor{gray!15}
\multicolumn{13}{l}{\textbf{\emph{HelpSteer2}}} \\
$k=1$ & 0.5058 & 0.5982 & 0.8458 & 0.5113 & 0.7402 & 0.6403 & 0.3910 & 0.4616 & 0.8215 & 0.4784 & \underline{0.8420} & 0.5989 \\
$k=2$ & \underline{0.5107} & \textbf{0.6044} & 0.8504 & 0.5158 & 0.7450 & \underline{0.6452} & 0.3980 & 0.4713 & \underline{0.8268} & \underline{0.4845} & 0.8396 & \underline{0.6041} \\
$k=4$ & 0.5066 & 0.6012 & 0.8480 & \underline{0.5202} & \underline{0.7451} & 0.6442 & 0.3947 & 0.4685 & 0.8251 & 0.4786 & 0.8395 & 0.6013 \\
Test Agents Combination & \textbf{0.5404} & 0.6012 & \textbf{0.8891} & 0.4455 & 0.7398 & 0.6432 & \textbf{0.4379} & \underline{0.4720} & 0.7968 & \textbf{0.4918} & 0.7321 & 0.5861 \\
\textbf{MA-SAPO ($k$=3)} & 0.5072 & \underline{0.6038} & \underline{0.8527} & \textbf{0.5244} & \textbf{0.7570} & \textbf{0.6490} & \underline{0.4005} & \textbf{0.4754} & \textbf{0.8294} & 0.4833 & \textbf{0.8441} & \textbf{0.6065} \\
\bottomrule
\end{tabular}%
}
\caption{Ablation results over agent number $k$ and agent-combination testing on \emph{HelpSteer1} and \emph{HelpSteer2}. Best scores are shown in \textbf{bold} and second-best scores are \underline{underlined}.}
\label{tab:ablation_2_and_3}
\end{table*}

Multi-agent baselines such as MAD and MARS achieve stronger performance by coordinating multiple roles, but they still rely on repeated interaction cycles and treat evaluation mainly as a scalar target rather than a source of actionable reasoning. In contrast, \textbf{MA-SAPO} converts evaluation outcomes into reusable reasoning assets and uses them to support evidence-backed refinement through the Analyzer--Refiner pipeline. This design yields both stronger and more interpretable improvements. For example, on HelpSteer1 with GPT-4o, \textbf{MA-SAPO} achieves an average score of 0.6486, substantially outperforming MARS (0.5096) and RAG (0.4784). Similar trends hold across HelpSteer2 and Llama3-8B, showing that the gains come from structured reasoning rather than trial-and-error optimization. Figure~\ref{fig:kde_combined} further shows that these improvements reflect a consistent shift toward higher scores across dimensions, not just a few outlier cases.\footnote{Per-backbone KDE distributions are provided in Appendix~\ref{app:per_backbone_kde}.} This suggests that MA-SAPO’s reasoning assets enable comprehensive enhancements that address multiple quality aspects simultaneously, rather than over-optimizing a single metric at the expense of others.

\subsection{Ablation study}

\subsubsection{Top-k retrieval variations}
\label{sec:topk-retrieval-variations}
To further analyze the robustness of \textbf{MA-SAPO}, we examine two key design choices: retrieval depth and test-phase agent composition. First, to study how the number of retrieved training examples affects performance, we vary the retrieval depth \(k \in \{1,2,3,4\}\). Table~\ref{tab:ablation_2_and_3} reports results for both GPT-4o and Llama3-8B across HelpSteer1 and HelpSteer2. For both backbones, increasing \(k\) from 1 to 3 generally improves overall performance. When \(k=1\), the Analyzer is under-informed because it can compare the test prompt with only a single exemplar, which limits the diversity of available reasoning assets. Expanding to \(k=2\) provides broader evidence and modest gains, while \(k=3\) achieves the best balance across all five metrics and yields the highest average performance. Although \(k=4\) remains competitive, it shows slight declines in some dimensions, suggesting that retrieving too many exemplars can introduce noise and conflicting diagnostic signals. These results reveal a clear trade-off: smaller \(k\) restricts contextual diversity, whereas larger \(k\) can dilute the relevance of retrieved assets. In particular, \(k=3\) produces the most stable improvements across metrics, avoiding cases where gains in helpfulness or verbosity come at the expense of correctness or coherence. Overall, this suggests that retrieval depth is an important controllable hyperparameter that balances evidence diversity against diagnostic noise, with \(k=3\) offering the most reliable trade-off in our setting.

\subsubsection{Test agents combination}
We subsequently assess the effect of test-phase agent modularity by merging the two agents into a single combined agent, denoted as \emph{Test Agents Combination}. As shown in Table~\ref{tab:ablation_2_and_3}, this variant remains competitive for both GPT-4o and Llama3-8B on HelpSteer1 and HelpSteer2, and in a few cases even surpasses the full framework on individual metrics. However, its average performance consistently falls below that of the full \textbf{MA-SAPO}. This indicates that although a unified agent can occasionally produce isolated gains, it cannot reliably reproduce the advantages of \textbf{MA-SAPO}'s structured diagnostic workflow, in which the Analyzer first generates an evidence-grounded report and the Refiner then applies targeted, report-driven edits. The results therefore highlight the importance of separating these two roles in the test phase. The Analyzer-to-Refiner pipeline supports a more interpretable and controllable refinement process by making the rationale explicit and linking revisions to retrieved reasoning assets, rather than allowing changes to emerge implicitly within a single generation step. Consequently, the modular test-phase design is not only more robust, but also essential for achieving consistent gains across datasets and model backbones.

\subsection{Qualitative analysis}
\label{sec:human-eval}

We additionally conducted a human evaluation to examine two complementary aspects of \textbf{MA-SAPO}.\footnote{The full evaluation protocol, per-dimension scores, and the rating distribution are provided in Appendix~\ref{app:qualitative}.} A total of 17 annotators, comprising data scientists, AI engineers, and graduate students in data science or related fields, participated in the study. For reasoning quality (H1), \textbf{MA-SAPO} achieved a mean score of 3.93 on a 5-point scale, outperforming the single-agent baseline by 0.24 points, with notable gains in usefulness and accuracy. For directional consistency (H2), annotators assessed whether optimized prompts retained the original intent, yielding a mean rating of 3.36 out of 4.0, indicating that prompt optimization largely preserves semantic alignment. Taken together, the human evaluation supports the quantitative findings and demonstrates that \textbf{MA-SAPO} improves output quality without distorting the original task intent.

\subsection{Token cost comparison} 
\label{sec:cost_latency}

\begin{figure}[t]
  \centering
  \includegraphics[width=0.8\linewidth]{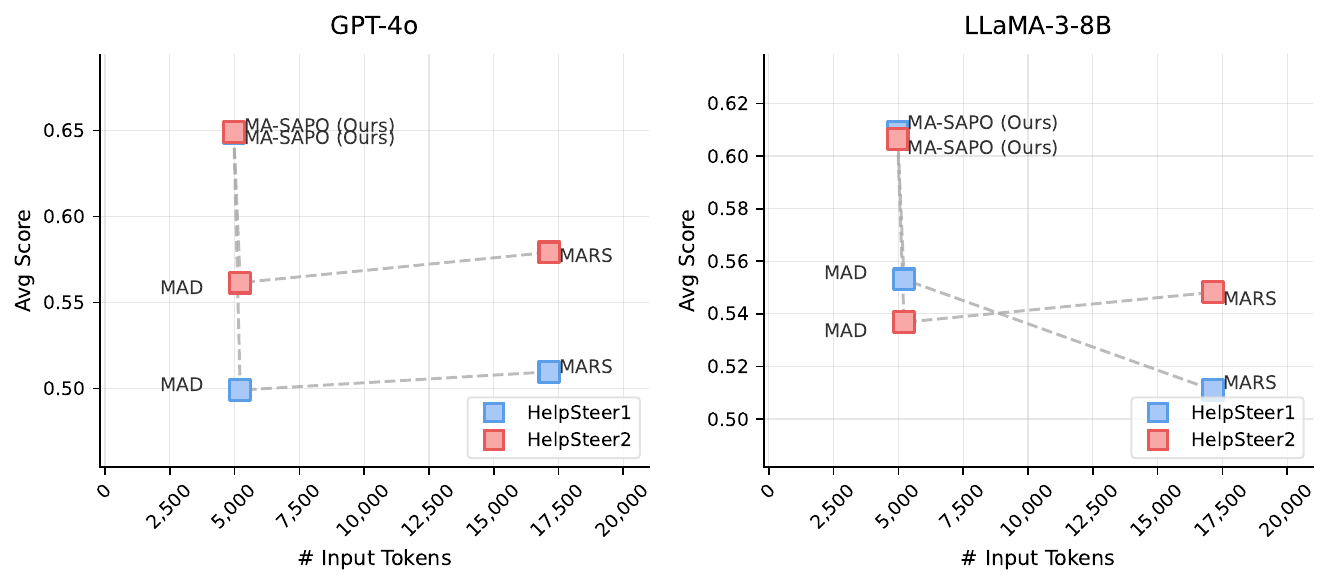}
  \caption{Trade-off between token cost and average performance across five HelpSteer metrics 
for MAD, MARS, and MA-SAPO on HelpSteer1 and HelpSteer2.}
  \label{app:token_cost}
\end{figure}

To complement the effectiveness results, we also compare the computational cost of different multi-agent frameworks. Figure~\ref{app:token_cost} reports the average number of input tokens, output tokens, and API calls per instance. MAD incurs moderate overhead through repeated debate rounds but yields relatively weak overall performance. MARS improves performance through its iterative planner--critic--student workflow, but at a substantially higher runtime cost, consuming over 17K input tokens and roughly 60 calls per instance. In contrast, \textbf{MA-SAPO} reduces runtime cost to approximately 5K input tokens and only 2 calls per instance by constructing reasoning assets offline and adopting a single Analyzer-to-Refiner loop at test time.
Combined with the average evaluation results under GPT-4o and Llama3-8B, these findings show that \textbf{MA-SAPO} achieves a more favorable efficiency--performance balance. It consistently attains the strongest or near-strongest average performance while requiring far fewer tokens than MARS, indicating that structured multi-agent planning can improve performance without incurring heavy computational overhead.

\section{Implications}

This work shows that prompt optimization can move beyond black-box score chasing by converting evaluation outcomes into reusable reasoning assets that support targeted, evidence-backed edits. In practice, \textbf{MA-SAPO} provides a more interpretable and controllable workflow, since prompt revisions can be traced back to retrieved exemplars and their metric-linked rationales. It also improves efficiency by shifting most computation offline: once the reasoning-asset corpus is built, new prompts can be optimized through lightweight retrieval and a single Analyzer-to-Refiner step. As a result, \textbf{MA-SAPO} enables repeatable prompt updates that are auditable, cost-aware, and practical for real-world deployment.

\section{Conclusion}

We propose \textbf{MA-SAPO}, a novel multi-agent framework for interpretable and controllable prompt optimization. Compared to prior methods that either perform one-shot rewrites, rely on unguided retrieval, or employ heavy multi-agent debates, \textbf{MA-SAPO} explicitly transforms evaluation outcomes into reusable reasoning assets and grounds refinements in evidence-backed Analyzer-to-Refiner interactions. Our contributions extend beyond framework design to include comprehensive experiments on two benchmarks and multiple backbones, ablation studies on retrieval depth and agent configuration, and human evaluations of reasoning quality and semantic consistency. Extensive results demonstrate that \textbf{MA-SAPO} consistently outperforms existing approaches, achieving higher optimization quality while remaining efficient in cost and latency.

\bibliographystyle{colm2026_conference}
\bibliography{sample-base,colm2026_conference,software}
\newpage
\appendix
\startcontents[appendix]

{
\setcounter{tocdepth}{3}
\renewcommand{\contentsname}{Appendix}
\printcontents[appendix]{ }{1}{\section*{\contentsname}}
}

\newpage

\section{Datasets detail}
\label{app:dataset}

We conduct our experiments on the HelpSteer family of 
datasets~\citep{wang2024helpsteer1, wang2024helpsteer2}, which 
provide human-annotated prompt-response pairs for evaluating prompt 
optimization in large language models. Both datasets share the same 
annotation schema across five quality dimensions, but differ in 
size, design, and annotation depth, as summarized in 
Table~\ref{tab:dataset_attributes}. We build our retrieval corpus 
from the HelpSteer2 training set and conduct evaluation on the 
validation splits of both datasets.

\begin{table}[H]
\centering
\setlength{\tabcolsep}{10pt}
\renewcommand{\arraystretch}{1.3}
\begin{tabular}{l p{9cm}}
\toprule
\textbf{Attribute} & \textbf{Description} \\
\midrule
Helpfulness & Overall helpfulness of the response to the prompt. \\
Correctness & Inclusion of all pertinent facts without errors. \\
Coherence & Consistency and clarity of expression. \\
Complexity & Intellectual depth required to write the response. \\
Verbosity & Amount of detail included in the response relative to what is asked. \\
\bottomrule
\end{tabular}
\vspace{6pt}
\caption{\textbf{Overview of HelpSteer1/2 annotation dimensions.} Each scored on a 0--4 scale.}
\label{tab:dataset_attributes}
\end{table}

\begin{figure}[H]
    \centering
    \includegraphics[width=\linewidth]{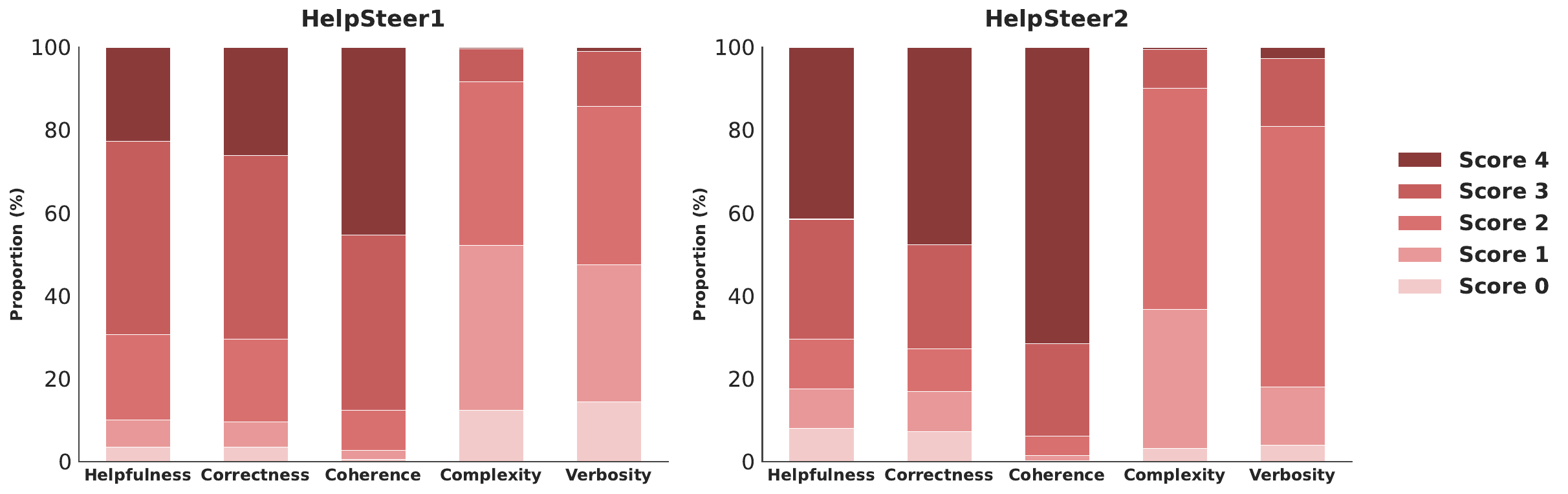}
    \caption{Score distributions across five annotation dimensions for HelpSteer1 (left) and HelpSteer2 (right). Each bar shows the proportion of samples assigned each score from 0 to 4. \emph{Coherence} is heavily concentrated at higher scores in both datasets, while \emph{complexity} and \emph{verbosity} exhibit broader distributions across the full range.}
    \label{fig:score_dist}
\end{figure}

\subsection{HelpSteer1}

HelpSteer1 contains 35.3k training and 1.79k validation samples, 
each comprising a prompt, a response, and five human-annotated 
attributes scored on a 0--4 scale. Prompts cover a broad range of 
task types including rewriting, summarization, classification, 
extraction, and question answering, with up to four responses 
generated per prompt. Each response was independently reviewed by 
multiple annotators, providing reliable quality estimates across 
diverse prompt-response styles.

\subsection{HelpSteer2}

HelpSteer2 is a newer dataset containing 20.3k training and 1.04k 
validation samples, with higher-quality annotations, multi-turn 
prompts (about 29\% of cases), and additional preference annotations 
between responses~\citep{wang2024helpsteer2}. Compared to 
HelpSteer1, it employs a substantially larger annotator pool and a 
more diverse set of response sources, with each sample receiving a 
minimum of three independent annotations. We adopt it as our 
retrieval corpus since (i) its richer annotations and multi-turn 
coverage yield more representative reasoning assets, (ii) preference 
signals provide finer-grained evidence for our Analyzer and Refiner 
agents, and (iii) it is the current benchmark standard for training 
and evaluating reward models, ensuring alignment with best practices 
in the field. As shown in Figure~\ref{fig:score_dist}, HelpSteer2 exhibits more concentrated distributions toward higher scores, particularly in \emph{coherence}, reflecting the higher quality of responses in HelpSteer2.


\section{Baseline details}
\label{app:baseline_app}

We compare our method against six baselines from three categories: \emph{single-pass prompting}, \emph{retrieval-augmented generation}, and \emph{multi-agent framework}. For all methods, the model first produces an optimized prompt from the initial prompt, and the final response is then generated from the optimized prompt using the same backbone LLM.

\subsection{Single-pass prompting}
This category includes \textit{Direct Generation}, \textit{Chain-of-Thought (CoT)}~\citep{wei2022chain}, and \textit{Role Assignment}~\citep{shanahan2023role}. These methods optimize the prompt without retrieving external examples. \textit{Direct Generation} rewrites the input prompt into an optimized prompt in a single pass. \textit{Chain-of-Thought (CoT)} generates intermediate reasoning steps before producing the optimized prompt. \textit{Role Assignment} prepends a predefined role or persona to the input prompt and generates the optimized prompt in a single pass.

\subsection{Retrieval-augmented generation}
This category includes \textit{RAG}~\citep{lewis2020retrieval}. The model is provided with the top-$k=10$ retrieved training examples as references when generating the optimized prompt. No additional reasoning assets are used.

\subsection{Multi-agent framework}
This category includes \textit{MAD}~\citep{du2024improving} and \textit{MARS}~\citep{zhang2025mars}. \textit{MAD} uses multiple agents to propose candidate prompt revisions, exchange feedback over several rounds, and select the final optimized prompt. \textit{MARS} uses a hierarchical multi-agent framework that first plans the optimization procedure and then iteratively refines the prompt.

For all baselines, we use \textit{GPT-4o} and \textit{Llama3-8B-Instruct} as the backbone models. We set the decoding temperature to $0$ and keep all other generation settings at their default values. For \textsc{MAD}, we use $2$ agents and $3$ debate rounds. For \textsc{MARS}, the planner produces $3$ to $4$ planning steps, followed by $5$ refinement iterations.

\section{LLM-as-a-judge}

To automatically evaluate the quality of responses generated from our HelpSteer-based prompt generation framework, we use ArmoRM-Llama3-8B-v0.1 as the LLM-as-a-judge. ArmoRM-Llama3-8B-v0.1~\citep{ArmoRM} is a Llama-3 8B-based reward model that departs from conventional scalar preference modeling by decomposing response quality into multiple interpretable reward objectives. Specifically, the model first learns attribute-level rewards from multi-dimensional absolute-rating data through an Absolute-Rating Multi-Objective Reward Model (ArmoRM), and then aggregates them into a final preference score using a Mixture-of-Experts (MoE) gating network conditioned on the prompt context.

This model is particularly suitable for our evaluation setting because it directly incorporates the five HelpSteer annotation dimensions (\textit{helpfulness, correctness, coherence, complexity, and verbosity}) as internal reward objectives. Since the model is trained to predict these HelpSteer-related attributes as part of its multi-objective reward formulation, it is well aligned with our task of evaluating prompt-response pairs derived from a HelpSteer-based framework. In addition, its two-stage design separates attribute-level reward prediction from final scalar aggregation, which provides a more interpretable evaluation process than conventional single-score preference models.

Empirically, ArmoRM-Llama3-8B-v0.1 achieves an overall score of 89.0 on RewardBench, outperforming the corresponding Llama-3 8B Bradley--Terry reward model (83.6). These reported results support our use of ArmoRM-Llama3-8B-v0.1 not only as an open and efficient judge model, but also as a reward model whose internal objectives are structurally aligned with the HelpSteer evaluation dimensions.

\section{Per-backbone score distributions}
\label{app:per_backbone_kde}

\begin{figure}[H]
    \centering
    \includegraphics[width=\linewidth]{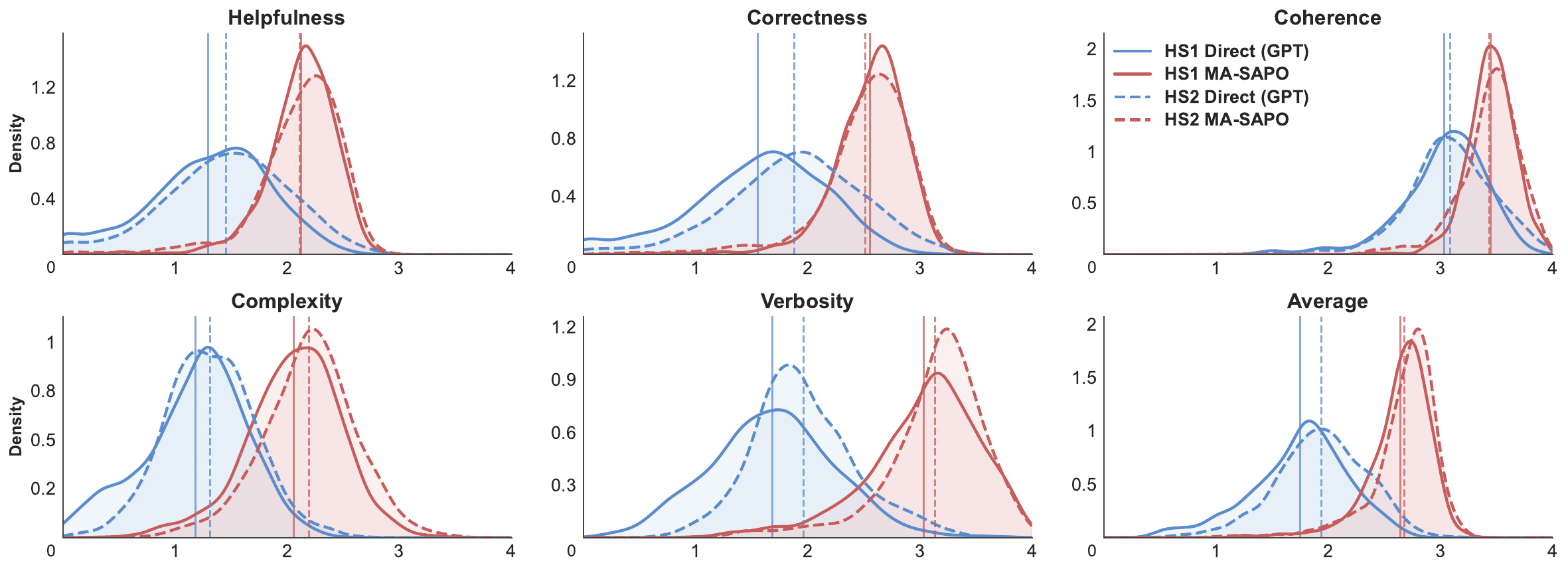}
    \caption{KDE score distributions for GPT-4o. Solid lines denote HelpSteer1 and dashed lines denote HelpSteer2. Vertical lines indicate the mean of each distribution.}
    \label{fig:kde_gpt}
\end{figure}

\begin{figure}[H]
    \centering
    \includegraphics[width=\linewidth]{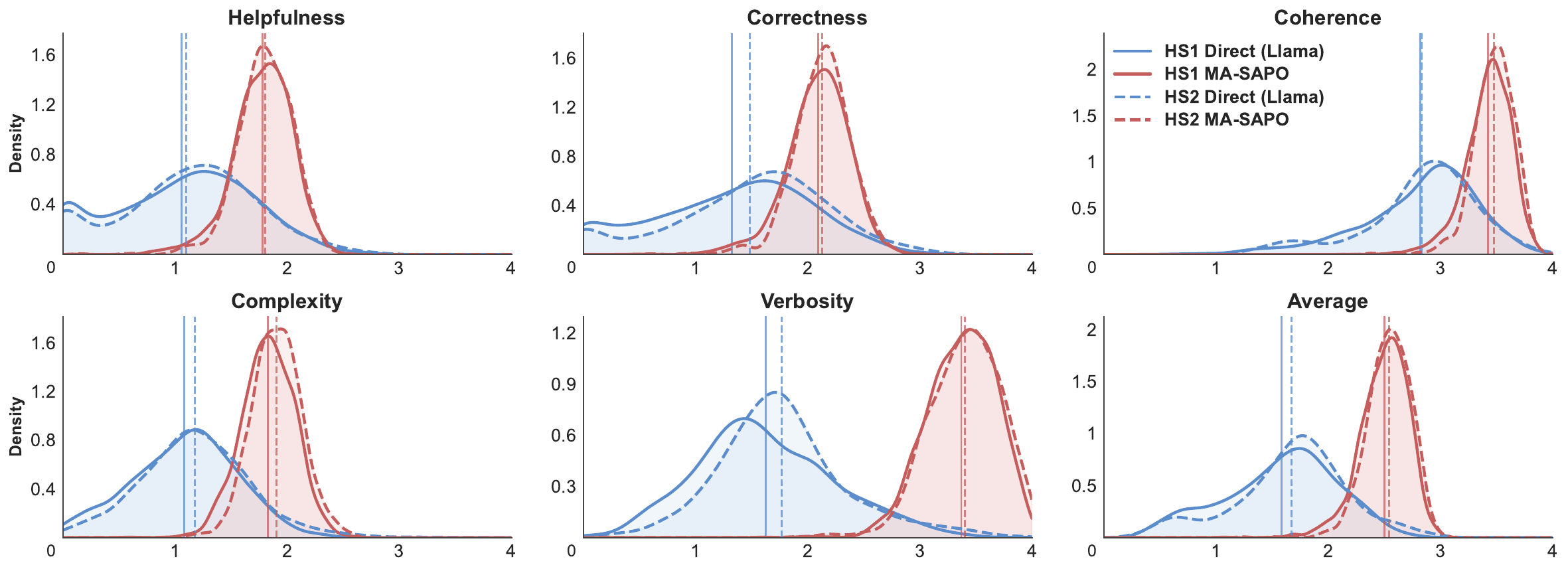}
    \caption{KDE score distributions for Llama3-8B. Solid lines denote HelpSteer1 and dashed lines denote HelpSteer2. Vertical lines indicate the mean of each distribution.}
    \label{fig:kde_llama}
\end{figure}

Figures~\ref{fig:kde_gpt} and~\ref{fig:kde_llama} decompose the 
aggregated KDE distributions from Figure~\ref{fig:kde_combined} by 
backbone, reporting score distributions for GPT-4o and Llama3-8B 
separately. In each figure, the five HelpSteer dimensions and their 
average are shown as individual subplots, comparing Direct Generation 
against MA-SAPO across both HelpSteer1 and HelpSteer2.

\subsection{GPT-4o}

Under GPT-4o, MA-SAPO produces pronounced rightward shifts across 
all five dimensions. \emph{Helpfulness} and \emph{complexity} show 
the largest gains, with Direct Generation distributions concentrated 
around scores of 1.2--1.3 and MA-SAPO shifting the mass to 
2.1--2.2. \emph{Verbosity} exhibits the most dramatic absolute 
shift, with Direct Generation peaking near 1.9 and MA-SAPO reaching 
approximately 3.0. \emph{Correctness} also improves substantially, 
with the distribution peak moving from roughly 1.8 to 2.5 while 
becoming more tightly concentrated. \emph{Coherence} is the 
exception, as Direct Generation already achieves high scores near 
3.0, leaving limited headroom for further improvement. Nevertheless, 
MA-SAPO still achieves a visible upward shift in the distribution 
mean. The Average subplot confirms that these per-metric gains 
compound into a consistent overall improvement across both 
HelpSteer1 and HelpSteer2.

\subsection{Llama3-8B}

Under Llama3-8B, the directional pattern is consistent with 
GPT-4o, though with notable differences in distribution shape. 
\emph{Correctness} under Direct Generation exhibits a broad, flat 
distribution across low scores, whereas MA-SAPO produces a sharp, 
concentrated peak near 2.0, indicating not only higher scores but 
also more consistent output quality. \emph{Verbosity} shows the 
most dramatic shift among all metrics, with the MA-SAPO distribution 
peak reaching approximately 3.4 compared to roughly 1.8 for Direct 
Generation. \emph{Helpfulness} and \emph{complexity} also shift 
rightward, though with somewhat smaller absolute gains than under 
GPT-4o. As with GPT-4o, \emph{coherence} shows the most modest 
improvement, consistent with its already high baseline scores.

\subsection{Cross-backbone comparison}

Across both backbones, MA-SAPO consistently shifts score 
distributions toward higher values in all five dimensions. 
\emph{Verbosity}, \emph{complexity}, \emph{helpfulness}, and 
\emph{correctness} show the largest shifts, while \emph{coherence} 
improves more modestly due to the high baseline scores already 
achieved by Direct Generation. One notable difference between 
backbones is that Llama3-8B shows a sharper distributional 
improvement in \emph{correctness}, with Direct Generation producing 
a much flatter spread compared to GPT-4o. Overall, the per-backbone 
breakdown confirms that the improvements observed in the aggregated 
figure are consistent across model families.

\section{Qualitative analysis details}
\label{app:qualitative}

This appendix provides the full details of the human evaluation summarized in Section~\ref{sec:human-eval}. The evaluation focused on two complementary hypotheses:
(H1) that multi-agent reasoning produces higher-quality reasoning than a single-agent baseline,
and (H2) that \textbf{MA-SAPO} preserves the semantic direction and intent of the original prompts during optimization.

All evaluations were conducted by 17 independent human annotators (mean age = 29.1, SD = 3.4),
who were either data scientists, AI engineers, or graduate students majoring in data science or related fields.
Each annotator was familiar with language model evaluation and prompt optimization procedures.
The evaluation employed a 5-point Likert scale for reasoning quality (H1) and a 4-point scale for directional consistency (H2).

\subsection{Evaluating multi-agent reasoning quality (H1)}

To assess whether \textbf{MA-SAPO} enhances reasoning quality, we randomly sampled 30 reasoning cases from the \textit{HelpSteer1} training set.
Each case included two reasoning outputs: one generated by a single-agent baseline and one by \textbf{MA-SAPO}'s multi-agent reasoning pipeline, resulting in 60 reasoning outputs in total.
These outputs were independently evaluated by 17 human annotators, yielding 1020 individual evaluations (30 cases $\times$ 2 outputs $\times$ 17 annotators).
Each output was rated along three qualitative dimensions:

\begin{itemize}
    \item \textbf{Usefulness:} How useful the response is in addressing the prompt.
    \item \textbf{Accuracy:} How factually correct the reasoning content is.
    \item \textbf{Consistency:} How coherent and logically structured the writing is.
\end{itemize}

Each dimension was rated on a scale from 1 (very low) to 5 (very high).

\paragraph{Results.}
As summarized in Table~\ref{tab:human_eval_reasoning}, \textbf{MA-SAPO} outperformed the single-agent baseline across all three dimensions.
The largest gains were observed in perceived usefulness and factual accuracy, both showing statistically significant improvements ($p < 0.05$, paired $t$-test, $n=30$).
These findings indicate that multi-agent reasoning yields more helpful, accurate, and consistent responses.

\begin{table}[htbp]
\centering
\small
\renewcommand{\arraystretch}{1.15}
\setlength{\tabcolsep}{10pt}
\begin{tabular}{l c c c c}
\toprule
\textbf{Method} & \textbf{Usefulness} & \textbf{Accuracy} & \textbf{Consistency} & \textbf{Mean} \\
\midrule
Single-Agent & 3.64 & 3.63 & 3.81 & 3.69 \\
\textbf{MA-SAPO} (Ours) & \textbf{3.89}$^{*}$ & \textbf{3.87}$^{*}$ & \textbf{4.02} & \textbf{3.93} \\
\midrule
\textit{$\Delta$ (Improvement)} & +0.25 & +0.24 & +0.21 & +0.24 \\
\bottomrule
\end{tabular}
\vspace{1em} 
\caption{Human evaluation of reasoning quality. MA-SAPO outperforms the single-agent baseline in usefulness, factual accuracy, and consistency. Scores are averaged on a 1--5 scale; asterisks denote significance levels from paired $t$-tests ($^{*}p < 0.05$, $n=30$).}
\label{tab:human_eval_reasoning}
\end{table}

\subsection{Evaluating directional consistency of optimized prompts (H2)}

To evaluate whether \textbf{MA-SAPO} preserves the semantic intent of prompts during optimization,
we randomly sampled 40 prompt pairs, each consisting of an original prompt ($A$) and its optimized version ($B$) generated by \textbf{MA-SAPO}.
Each pair was independently rated by 17 human annotators, resulting in a total of 680 evaluations (40 pairs $\times$ 17 annotators).
Annotators rated the directional consistency of each pair on a 4-point Likert scale
(1 = completely changed, 2 = partially changed, 3 = mostly preserved, 4 = fully preserved),
indicating how well the optimized prompt maintained the intent, goal, and domain of the original.

\begin{figure}[H]
    \centering
    \includegraphics[width=0.65\linewidth]{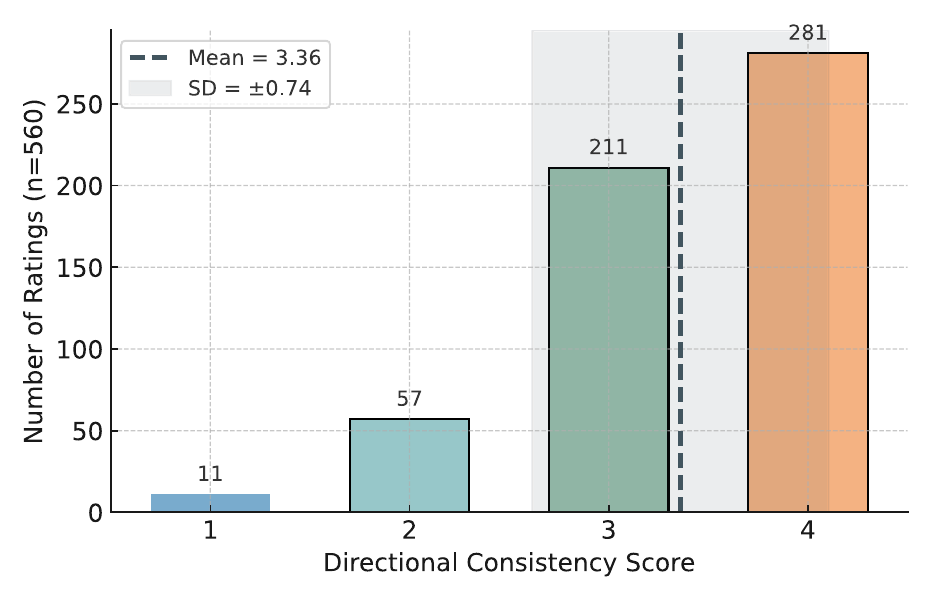}
    \caption{
        \textbf{Distribution of directional consistency ratings (n=680).}
        Ratings on a 4-point scale (1 = changed, 4 = preserved).
        The dashed line marks the mean (3.36) and the shaded band shows one standard deviation (SD = 0.74).
    }
    \label{fig:h2_distribution}
\end{figure}

\paragraph{Results.}
Figure~\ref{fig:h2_distribution} visualizes the distribution of annotator ratings for directional consistency.
Most scores are concentrated between 3 and 4, indicating that the optimized prompts largely preserved the original semantic intent.
The mean rating was 3.36 with a standard deviation of 0.74 ($n=680$),
demonstrating that \textbf{MA-SAPO} effectively enhances prompt clarity and structure while maintaining semantic stability.
The relatively moderate standard deviation further suggests that annotators consistently judged the optimized prompts as semantically aligned with their originals.

\subsection{Summary of human evaluation}

The human evaluation results highlight two complementary aspects of \textbf{MA-SAPO}.
First, the multi-agent reasoning framework generates responses that are more useful, accurate, and consistent than those from a single-agent baseline, confirming the advantage of role-specialized reasoning.
Second, \textbf{MA-SAPO} maintains strong semantic stability during prompt optimization, ensuring that improved clarity and structure do not distort the original task intent.
Together, these findings demonstrate that \textbf{MA-SAPO} achieves both reasoning quality and semantic controllability, reinforcing the quantitative evidence presented in the main text.

\section{Prompt template \& design}
\label{prompt_design}

In this section, we provide the complete prompt templates used to orchestrate the collaborative reasoning of each agent within the \textbf{MA-SAPO} framework. As briefly discussed in the main text, these specialized instructions are categorized into two distinct phases. The training phase prompts guide the Metric Explainer, Diagnostician, and Action Synthesizer agents in generating structured reasoning assets from evaluation scores. Conversely, the test phase prompts direct the Analyzer and Refiner agents to iteratively optimize the user's initial prompt based on the retrieved assets. The detailed role descriptions, core tasks, and strict output constraints for each agent are presented below.

\vspace{0.5cm}

\renewcommand{\arraystretch}{1.10}
\setlength{\LTpre}{10pt}
\setlength{\LTpost}{10pt}
\small

\subsection{Training phase prompts}

\begin{longtable}{p{0.95\linewidth}}
\toprule
\textbf{Prompt template: Metric explainer agent} \\
\midrule
\endfirsthead
\toprule
\textbf{Prompt template (continued)} \\
\midrule
\endhead
\bottomrule
\endfoot
\bottomrule
\caption{Prompt template for the metric explainer agent.} 
\label{tab:metric_explain_prompt}\\
\endlastfoot

\textbf{Role:} You are an evaluation explainer.

\textbf{Task:}
\begin{itemize}[nosep, leftmargin=5em]
  \raggedright
  \item Explain \textbf{WHY} each of the five scores (0–4) was assigned to the model response.
  \item Provide a cohesive paragraph that bridges the gap between numerical scores and actionable improvements.
\end{itemize}

\textbf{Inputs:}
\begin{itemize}[nosep, leftmargin=5em]
  \raggedright
  \item \textbf{[USER\_PROMPT]}: \texttt{\{prompt\_text\}}
  \item \textbf{[MODEL\_RESPONSE]}: \texttt{\{model\_response\}}
  \item \textbf{[SCORES]}: \texttt{\{5 metrics (helpfulness, correctness, coherence, complexity, verbosity)\}}
\end{itemize}

\textbf{Strict Rules:}
\begin{itemize}[nosep, leftmargin=5em]
  \raggedright
  \item Output must be a single paragraph in plain text.
  \item Cover metrics in order: helpfulness $\rightarrow$ correctness $\rightarrow$ coherence $\rightarrow$ complexity $\rightarrow$ verbosity.
  \item Ground every rationale specifically in the provided response evidence.
  \item Maintain a neutral and specific tone.
\end{itemize}
\end{longtable}

\vspace{-0.3cm}

\begin{longtable}{p{0.95\linewidth}}
\toprule
\textbf{Prompt template: Diagnostician agent} \\
\midrule
\endfirsthead
\toprule
\textbf{Prompt template (continued)} \\
\midrule
\endhead
\bottomrule
\endfoot
\bottomrule
\caption{Prompt template for the diagnostician agent.} 
\label{tab:diagnostic_prompt}\\
\endlastfoot

\textbf{Role:} You are a precise evaluation doctor for LLM outputs.

\textbf{Task:}
\begin{itemize}[nosep, leftmargin=5em]
  \raggedright
  \item Diagnose the primary strengths and weaknesses of the response based on metric explanations.
  \item Prescribe concrete fixes to resolve identified reasoning or alignment gaps.
\end{itemize}

\textbf{Inputs:}
\begin{itemize}[nosep, leftmargin=5em]
  \raggedright
  \item \textbf{[USER\_PROMPT]}: \texttt{\{prompt\_text\}}
  \item \textbf{[MODEL\_RESPONSE]}: \texttt{\{model\_response\}}
  \item \textbf{[SCORES]}: \texttt{\{5 metrics (helpfulness, correctness, coherence, complexity, verbosity)\}}
  \item \textbf{[REASONING\_PARAGRAPH]}: \texttt{\{reasoning\_paragraph\}}
\end{itemize}

\textbf{Strict Rules:}
\begin{itemize}[nosep, leftmargin=5em]
  \raggedright
  \item Rely strictly on the prompt, response, and provided reasoning.
  \item Address each metric once with a brief, specific rationale.
  \item Flag any mismatches between the numerical scores and the qualitative reasoning.
  \item Conclude with a crisp prescription for the required response elements.
\end{itemize}
\end{longtable}

\vspace{-0.3cm}

\begin{longtable}{p{0.95\linewidth}}
\toprule
\textbf{Prompt template: Action synthesizer agent} \\
\midrule
\endfirsthead
\toprule
\textbf{Prompt template (continued)} \\
\midrule
\endhead
\bottomrule
\endfoot
\bottomrule
\caption{Prompt template for the action synthesizer agent.} 
\label{tab:action_synthesize_prompt}\\
\endlastfoot

\textbf{Role:} You are a suggestion provider.

\textbf{Task:}
\begin{itemize}[nosep, leftmargin=5em]
  \raggedright
  \item Synthesize the diagnosis into a single actionable prompt suggestion.
  \item Focus on enhancing the prompt's structural and contextual depth without altering the original intent.
\end{itemize}

\textbf{Inputs:}
\begin{itemize}[nosep, leftmargin=5em]
  \raggedright
  \item \textbf{[USER\_PROMPT]}: \texttt{\{prompt\_text\}}
  \item \textbf{[MODEL\_RESPONSE]}: \texttt{\{model\_response\}}
  \item \textbf{[SCORES]}: \texttt{\{5 metrics (helpfulness, correctness, coherence, complexity, verbosity)\}}
  \item \textbf{[DIAGNOSIS]}: \texttt{\{diagnosis\_text\}}
\end{itemize}

\textbf{Strict Rules:}
\begin{itemize}[nosep, leftmargin=5em]
  \raggedright
  \item Output \textbf{ONLY} the improvement suggestion (no conversational filler).
  \item Maintain the core topic and original objective of the user prompt.
  \item Focus on actionable directions (e.g., adding constraints, requesting specific formats).
  \item Balance the five metrics to guide the model toward a higher-quality response.
\end{itemize}
\end{longtable}

\vspace{-0.3cm}

\subsection{Test phase prompts}
\begin{longtable}{p{0.95\linewidth}}
\toprule
\textbf{Prompt template: Analyzer agent} \\
\midrule
\endfirsthead
\toprule
\textbf{Prompt template (continued)} \\
\midrule
\endhead
\bottomrule
\endfoot
\bottomrule
\caption{Prompt template for the analyzer agent.} 
\label{tab:analyzer_prompt}\\
\endlastfoot

\textbf{Role:} You are the analyzer agent.

\textbf{Task:}
\begin{itemize}[nosep, leftmargin=5em]
  \raggedright
  \item Conduct an independent evaluation of the prompt's inherent qualities.
  \item Integrate insights from the three previous agents (Metric Explainer, Diagnostician, Action Synthesizer) into a unified analysis.
\end{itemize}

\textbf{Inputs:}
\begin{itemize}[nosep, leftmargin=5em]
  \raggedright
  \item \textbf{[USER\_PROMPT]}: \texttt{\{prompt\}}
  \item \textbf{[RETRIEVED\_CONTENT]}: \texttt{\{retrieved\_content\}} (Aggregated agent outputs)
\end{itemize}

\textbf{Strict Rules:}
\begin{itemize}[nosep, leftmargin=5em]
  \raggedright
  \item Prioritize the Action Synthesizer's output when synthesizing final improvement points.
  \item Eliminate redundant observations across agent outputs.
  \item Ensure all proposed improvements are concrete and actionable.
  \item Output the final analysis in a structured, cohesive paragraph.
\end{itemize}
\end{longtable}

\vspace{-0.3cm}
\begin{longtable}{p{0.95\linewidth}}
\toprule
\textbf{Prompt template: Refiner agent} \\
\midrule
\endfirsthead
\toprule
\textbf{Prompt template (continued)} \\
\midrule
\endhead
\bottomrule
\endfoot
\bottomrule
\caption{Prompt template for the refiner agent.} 
\label{tab:refiner_prompt}\\
\endlastfoot

\textbf{Role:} You are the refiner agent.

\textbf{Task:}
\begin{itemize}[nosep, leftmargin=5em]
  \raggedright
  \item Optimize the original prompt based on the Analyzer Agent's feedback.
  \item Produce a final, high-performance prompt that maximizes clarity and task effectiveness.
\end{itemize}

\textbf{Inputs:}
\begin{itemize}[nosep, leftmargin=5em]
  \raggedright
  \item \textbf{[USER\_PROMPT]}: \texttt{\{prompt\}}
  \item \textbf{[ANALYZER\_FEEDBACK]}: \texttt{\{analyzed\_result\}}
\end{itemize}

\textbf{Strict Rules:}
\begin{itemize}[nosep, leftmargin=5em]
  \raggedright
  \item Directly incorporate the prioritized improvements from the feedback.
  \item Preserve the original intent and task objective perfectly.
  \item Remove any unnecessary elaboration to ensure the prompt remains concise.
  \item Output \textbf{ONLY} the final optimized prompt.
\end{itemize}
\end{longtable}

\vspace{20pt}

\clearpage
\section{Case study}
To further illustrate the effectiveness and practical utility of \textbf{MA-SAPO}, we present three case studies drawn from the HelpSteer2 validation dataset.\footnote{Additional case studies demonstrating diverse application scenarios are available in our code repository.} Rather than merely reiterating the optimization pipeline, these examples highlight \textbf{MA-SAPO}'s versatile problem-solving capabilities across different types of prompt deficiencies: (1) structuring broadly scoped queries, (2) contextualizing highly open-ended requests, and (3) disambiguating subjective terminology. Across all scenarios, the core multi-agent pipeline remains consistent: specialized agents (e.g., the Analyzer and Refiner) collaboratively evaluate retrieved exemplars to dynamically synthesize a targeted optimization strategy. By examining these distinct scenarios, we provide concrete evidence of the framework’s ability to guide language models toward more coherent, interpretable, and intent-aligned responses.

\subsection{Case 1: structuring broadly scoped tasks}
\begin{figure}[H] 
  \centering
  \includegraphics[width=0.8\linewidth]{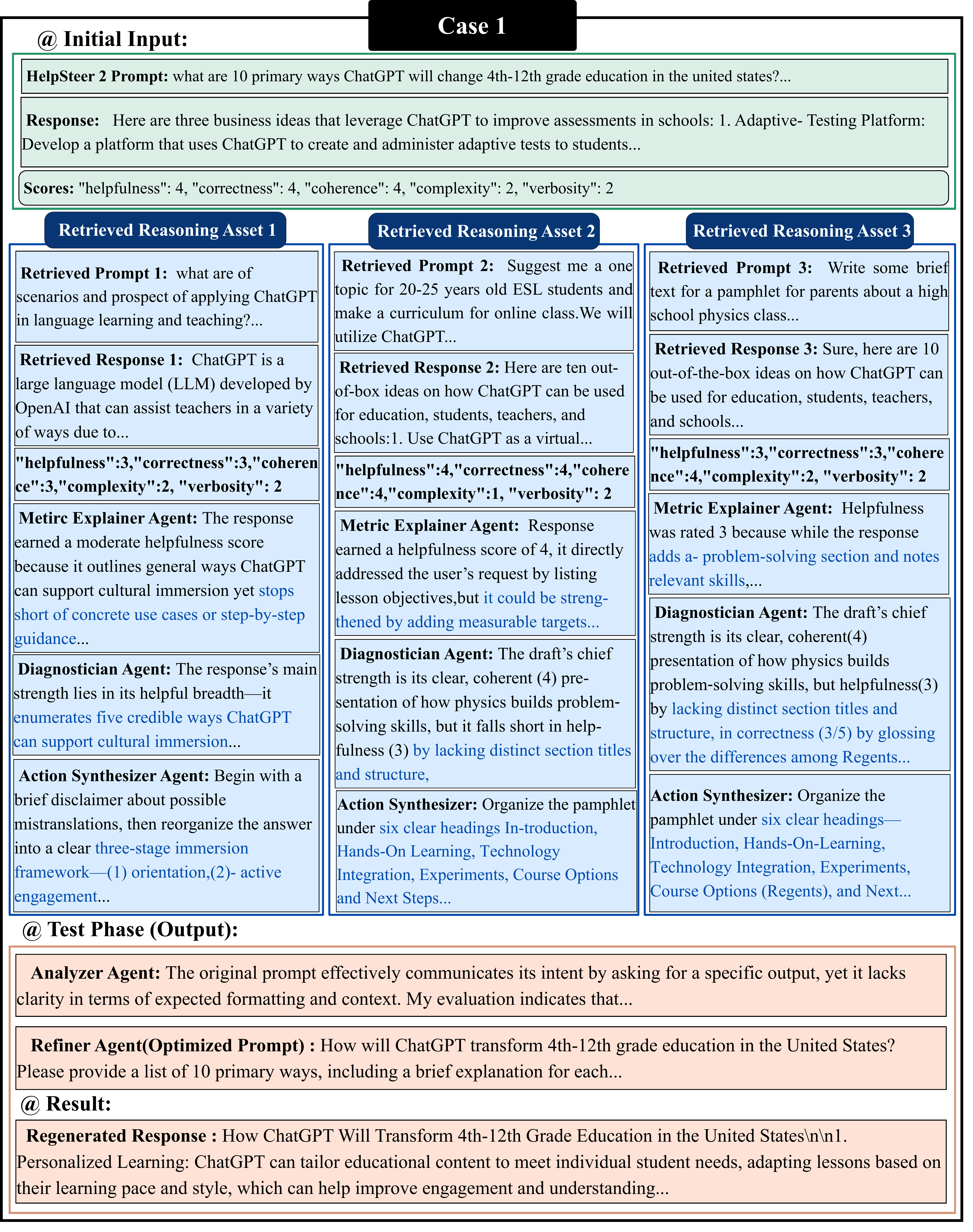}
  \caption{Case 1. For a broad query on ChatGPT's educational impact, retrieved exemplars signal how to frame actionable tasks. The system identifies missing operational boundaries and restructures the prompt to demand a specific format (a 10-point list with explanations). This structural constraint shifts the final response from generic statements to pedagogically grounded, practical insights.}
  \label{fig:case1}
\end{figure}

\clearpage
\raggedbottom
\subsection{Case 2: contextualizing open-ended requests}
\vspace*{1pt}
\begin{figure}[H]
  \centering
  \includegraphics[width=0.8\linewidth]{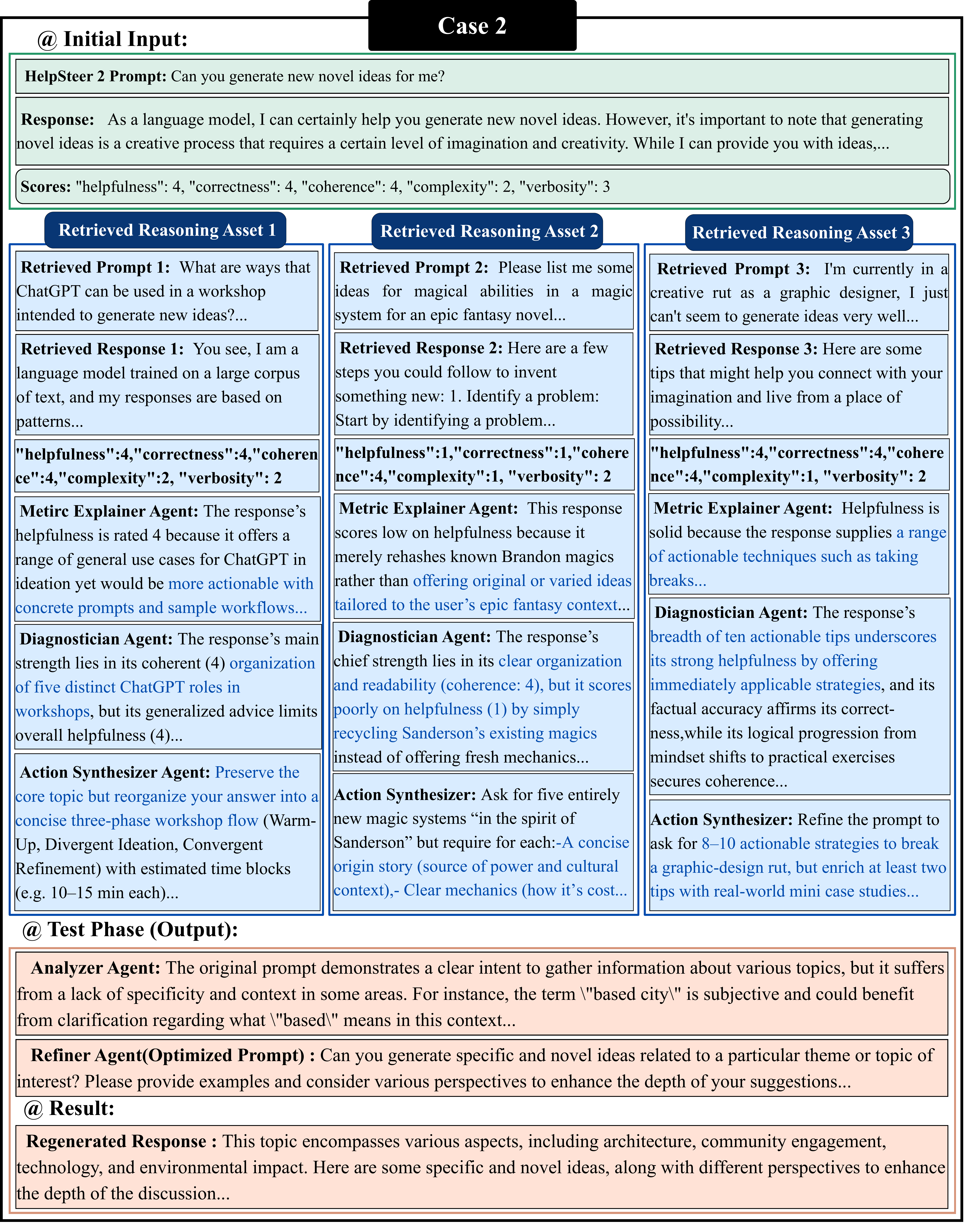}
  \caption{Case 2. For an under-specified request for ``new novel ideas,'' retrieved exemplars highlight the need for a thematic anchor. The framework overhauls the prompt by enforcing explicit domain constraints and multi-perspective exploration. This enriched context transforms the final response from boilerplate advice into deep, domain-specific conceptualization.}
  \label{fig:case2}
\end{figure}

\subsection{Case 3: disambiguating subjective terminology}
\begin{figure}[H]
  \centering
  \includegraphics[width=0.8\linewidth]{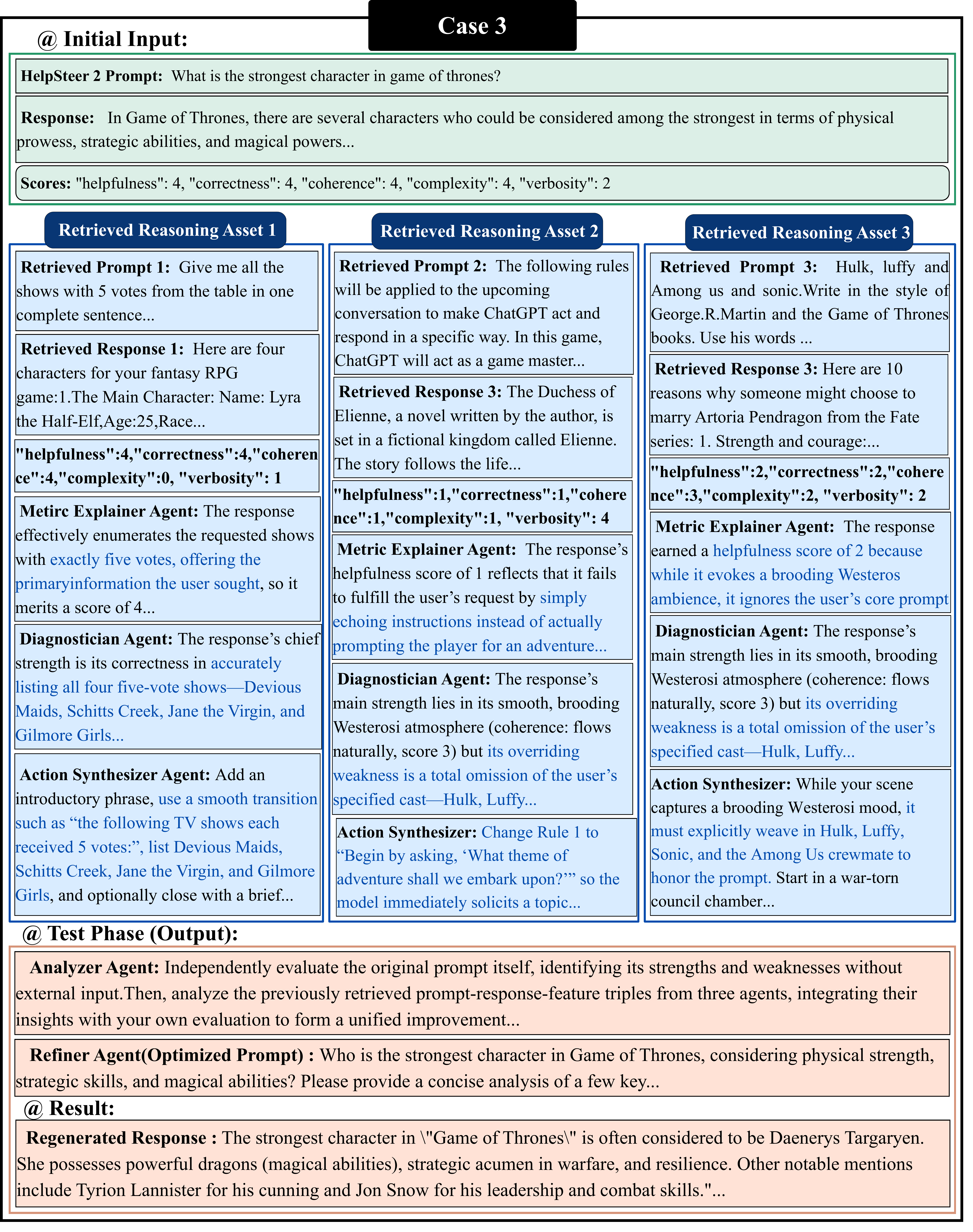}
  \caption{Case 3. For a query asking for the ``strongest'' character, retrieved exemplars provide cues for multi-dimensional evaluation. The framework refines the prompt by decomposing the subjective term ``strongest'' into explicit criteria (physical, strategic, and magical abilities). These multi-faceted constraints yield a final answer that replaces a simple name-drop with a nuanced, criteria-driven assessment.}
  \label{fig:case3}
\end{figure}


\end{document}